\documentclass[aip,jcp,reprint]{revtex4-1}
\usepackage{amsmath}
\usepackage{bm}
\usepackage{graphicx}
\usepackage{color}
\usepackage[utf8]{inputenc}

\newcommand{\kT}{k_{\mathrm{B}}T}
\renewcommand{\d}[1]{\mathrm{d}#1\,}

\begin{document}
	\title{Phase behavior of active Brownian disks, spheres, and dumbbells}
	\date{\today}
	\author{Jonathan Tammo Siebert}
	\affiliation{Institut für Physik, Johannes Gutenberg-Universität Mainz}
	\author{Janina Letz}
	\affiliation{Institut für Physik, Johannes Gutenberg-Universität Mainz}
	\author{Thomas Speck}
	\affiliation{Institut für Physik, Johannes Gutenberg-Universität Mainz}
	\author{Peter Virnau}
	\affiliation{Institut für Physik, Johannes Gutenberg-Universität Mainz}
	\begin{abstract}
		In this paper we provide high precision estimates of the phase diagram of active Brownian particles. We extract coexisting densities from simulations of phase separated states in an elongated box (slab geometry) which minimizes finite-size effects and allows for precise determination of points on the binodal lines. Using this method, we study the influence of both shape and dimensionality on the two-phase region. Active spheres and dumbbells of active particles are compared to the known phase diagram of active Brownian disks. In the case of dimers, both correlated and uncorrelated propulsion of the two beads are studied. The influence of correlation is discussed through a simple mapping.
	\end{abstract}
	\maketitle
	\section{Introduction\label{sec:introduction}}
	Active particles have been a major focus of research in soft matter and non-equilibrium physics in recent years. Their defining property, namely a directed self-propulsion, leads to a range of fascinating collective behavior for interacting particles. Early studies were mainly motivated by biological systems such as schools of fish, flocks of birds\cite{Vicsek:1995}, and swarming of bacteria\cite{Cates:2012}. Especially, natural microswimmers like algae\cite{Goldstein:2009,Friedrich:2012}, bacteria\cite{Wensink:2012a}, and sperm cells\cite{Alvarez:2012,Alvarez:2014,Magdanz:2013} are now studied both experimentally and by means of computer simulations.\par
	In the last decade, also artificial microswimmers propelled by a multitude of mechanisms have been produced. One class are rod-like swimmers\cite{Paxton:2006,Hong:2007} or Janus particles driven by catalytic reactions\cite{Palacci:2010} of fuel that is provided by the surrounding solvent. Alternatively, when illuminated by a laser Janus particles whose hemispheres absorb laser light differently can also be propelled by self-thermophoresis\cite{Jiang:2010}, or local demixing of a surrounding solvent that is just below its critical point\cite{Buttinoni:2013}. Other approaches include cells propelled by an artificial filament that is magnetically driven\cite{Dreyfus:2005} or liquid droplets propelled by Marangoni stress induced flow\cite{Peddireddy:2012,Herminghaus:2014}.\par
	Besides the exact swimming mechanism of individual swimmers, another major focus of research is the fascinating collective behavior of such swimmers, including examples like swarming\cite{Wensink:2012}, turbulent motion\cite{Dombrowski:2004}, giant number fluctuations\cite{Ramaswamy:2003,Narayan:2007}, and clustering\cite{Bialke:2012}. A very common and successful model system for self-propelled colloids are active Brownian particles\cite{Bialke:2015a,Fily:2012,Redner:2013,Solon:2015,Stenhammar:2014,Wysocki:2014}. This model is rather minimalistic. Nonetheless, it still shows a motility induced phase separation\cite{Tailleur:2008} which is found in many active systems\cite{Cates:2010,Zhang:2010,Theurkauff:2012,Palacci:2013,Buttinoni:2013,Zoettl:2014}. This separation closely resembles a gas-liquid transition in equilibrium\cite{Bialke:2015a} and thus can serve as a model system to study a non-equilibrium phase transition. Several studies have examined the nucleation kinetics\cite{Richard:2016} as well as the phase behavior by mean field analysis\cite{Speck:2015} or a Maxwell construction on the pressure\cite{Solon:2015}. The phase diagram in both two\cite{Redner:2013,Fily:2014} and three dimensional systems\cite{Stenhammar:2014,Wysocki:2014} was examined by computer simulations. A more quantitative analysis of the phase diagram taking finite size effects into account has been done for the two dimensional case\cite{Bialke:2015a}.\par
	Determination of the phase diagram by use of computer simulations requires careful treatment of effects of finite box sizes. As systems that can be treated numerically are always far from the thermodynamic limit, system size and boundary conditions have a large influence on the system. Simple scanning for nucleation or examination of stability of phase separation will strongly depend on both the time scale that is simulated as well as the system size that is studied. A reliable method proven to work in equilibrium\cite{Watanabe:2012} is to extract coexisting densities in a phase separated state. If studied in a slab configuration, this will yield the correct bulk densities already for comparably small systems. The slab configuration is needed to ensure straightness of the interface, as curved interfaces will lead to a Laplace pressure, that will influence the density. This concept still holds in our non-equilibrium system, in which also a positive line tension of the interface was observed\cite{Bialke:2015a}.\par
	While the original active Brownian particles are disk-like, and purely repulsive, also particles with attraction and their phase diagram have been studied\cite{Redner:2013}. Recently, even the influence of activity on the critical point has been determined\cite{Prymidis:2016}. Furthermore, non-spherical agents have been examined. Especially, elongated shapes as rod-like\cite{Peruani:2006,Yang:2010,Wensink:2012} or dumbbell swimmers\cite{Suma:2014,Tung:2016} and even longer chains of active particles\cite{Kaiser:2015} were considered. Here, the type of active propulsion seems to play an important role\cite{Cugliandolo:2015a}. In this study, the phase diagram of active Brownian monomers in three dimensions and dimers in two dimensions is evaluated, using a method similar to earlier work on the original active Brownian particles\cite{Bialke:2015a}. This allows us to study influences of dimensionality and anisotropic geometry in isolation.\par
	In the following, we show, that active Brownian dimers with both uncorrelated and correlated propulsion directions phase separate only for larger propulsion velocities than corresponding monomers by examination of the binodal lines in their phase diagrams. The differences between both propulsion mechanisms are discussed through a simple mapping. Furthermore, we present precise estimates for the binodal lines in a system of active Brownian spheres.
	\section{Model\label{sec:model}}
	\subsection{Monomers}
	We simulate particles interacting via a strongly repulsive WCA-potential that is cut off at $r_{\mathrm{cut}} = 2^{1/6}\sigma$:
	\begin{equation}
	U_{\mathrm{WCA}}(r_{ij}) = 4\epsilon\left(\left(\frac{\sigma}{r_{ij}}\right)^{12}-\left(\frac{\sigma}{r_{ij}}\right)^6+\frac{1}{4}\right)\text{.}
	\end{equation}
	Here $r_{ij}\equiv \left|\bm{r}_i-\bm{r}_j\right|$ is the distance between two particles and $\epsilon$ is a parameter for the steepness of the potential chosen to be $100\kT$. These particles can be mapped onto hard disks in two and spheres in three dimensions via a Barker-Henderson diameter\cite{Barker:1967} $d_{\mathrm{BH}} \approx 1.10688\sigma$. Generally, this mapping only works for equilibrium systems. Nonetheless, due to the steepness of the potential, deviations in the non-equilibrium case for self-propelled particles are neglected.\par
	Commonly, active Brownian particles are examined by solving the overdamped Langevin equation with an additional term governing the self-propulsion\cite{Bialke:2015a,Fily:2012,Redner:2013,Solon:2015,Stenhammar:2014,Wysocki:2014}:
	\begin{equation}
		\dot{\bm{r}}_i = -\frac{D}{\kT}\nabla U\left(\left\{\bm{r}_i\right\}\right) + \sqrt{2 D} \bm{R}_{\mathrm{t}} + v_0 \bm{e}_i\text{,}
	\end{equation}
	with uncorrelated Gaussian translational noise $\bm{R}_{\mathrm{t}}$ with zero mean and unit variance modeling the solvent.\par
	The propulsion is modeled as an additional velocity of constant magnitude $v_0$. Its direction undergoes free rotational diffusion with rotational diffusion coefficient $D_{\mathrm{r}} = \frac{3 D}{d_{\mathrm{BH}}^2}$, assuming no-slip boundary conditions. In two dimensions, the propulsion direction $\bm{e}_i$ follows as
	\begin{equation}
		\bm{e}_i = \begin{pmatrix}\cos\phi_i\\
		\sin\phi_i\end{pmatrix}\text{, with }
		\dot{\phi}_i = \sqrt{2 D_{\mathrm{r}}} R_{\mathrm{r}}\text{,}
	\end{equation}
	where $R_{\mathrm{r}}$ is an uncorrelated Gaussian white noise random variable with zero mean and unit variance, as well.\par
	In three dimensions, the propulsion direction's time evolution is given by\cite{Stenhammar:2014,Wysocki:2014}:
	\begin{equation}
		\dot{\bm{e}}_i = \sqrt{2 D_{\mathrm{r}}} \bm{e}_i \times\bm{R}_{\mathrm{r}}\text{.}
	\end{equation}
	For finite time steps, simple integration of this equation would lead to non-normalized orientation vectors. Thus, $\bm{e}_i$ is normalized after each time step.\par
	From here on, $\sigma$, $\sigma^2/D$, and $\kT$ are used as units of length, time, and energy respectively and will thus be omitted.
	\subsection{Dimer model}
	\begin{figure}[tb]
		\begin{minipage}{0.49\linewidth}
			\includegraphics[width=\linewidth]{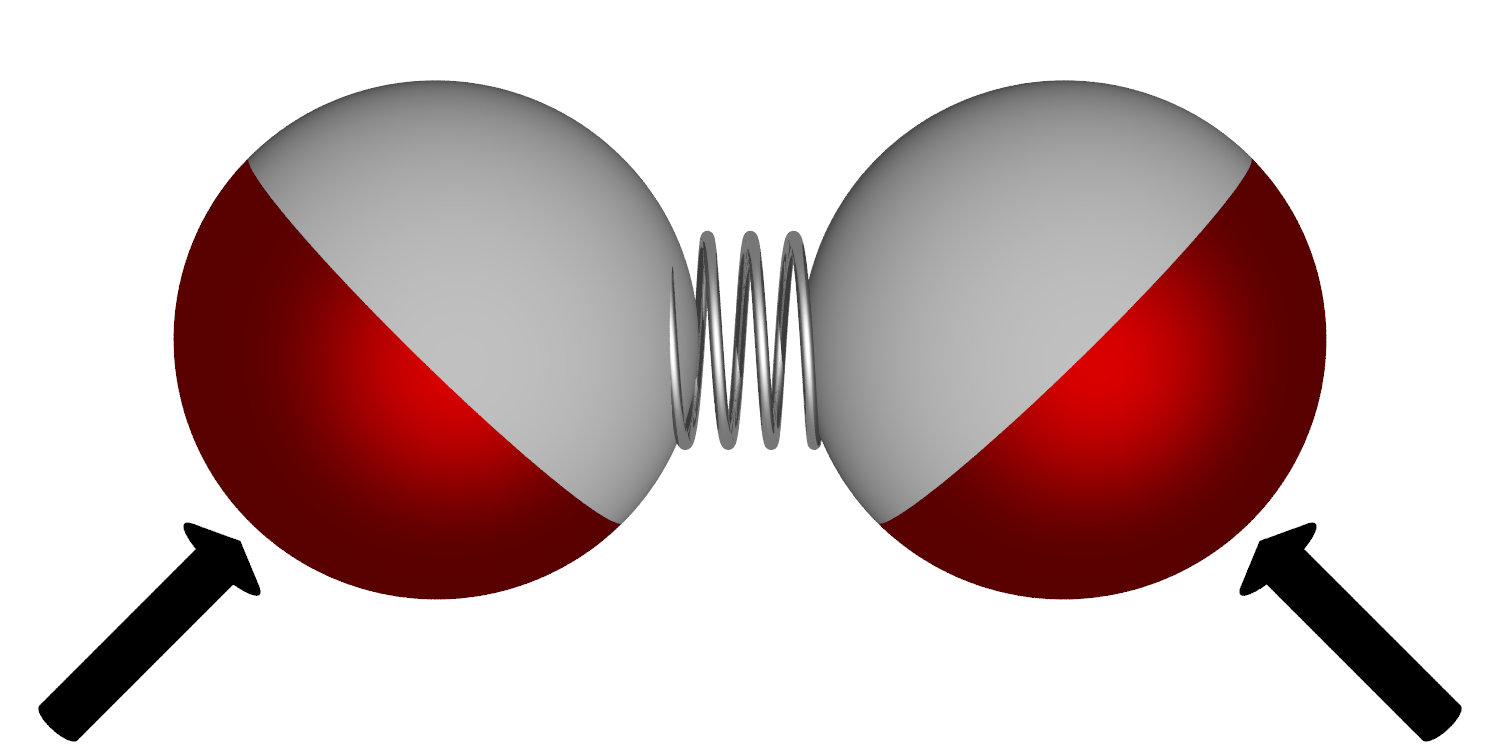}\\
			{\centering (a) uncorrelated}
		\end{minipage}
		\begin{minipage}{0.49\linewidth}
			\includegraphics[width=\linewidth]{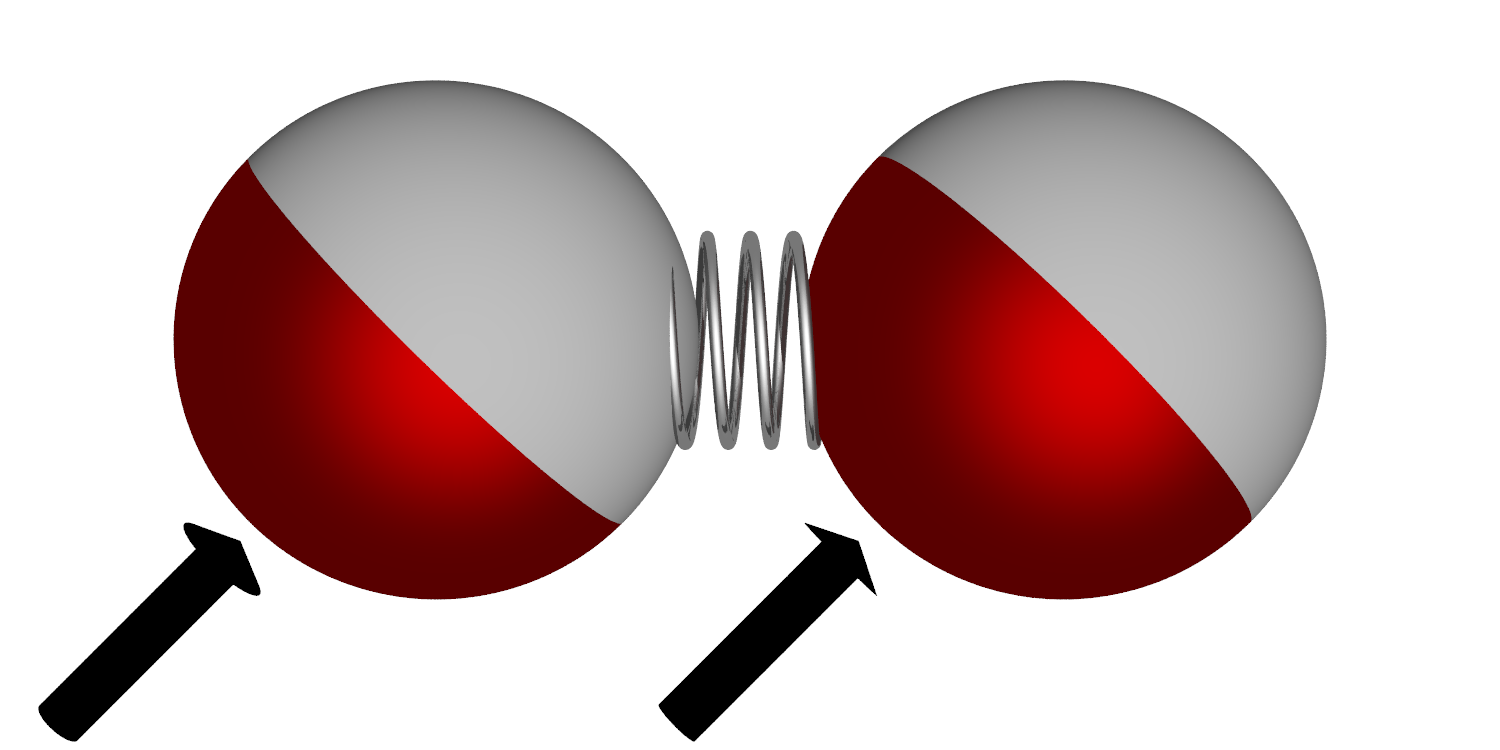}\\
			{\centering (b) correlated}
		\end{minipage}
		\caption{Schematic representation of both dimer types examined in this work: The propulsion direction is indicated by the black arrows. (a) Uncorrelated dimers undergo free rotational diffusion individually, resulting in unaligned propulsion directions. (b) Correlated dimers share a common director which undergoes free rotational diffusion. Therefore, their propulsion directions are always parallel.}
		\label{fig:schematicDimers}
	\end{figure}
	For active dimers\cite{Cugliandolo:2015}, pairs of particles are bonded by an additional FENE-potential:
	\begin{equation}
	U_{\mathrm{FENE}}(r_{ij}) = -\frac{K R_0^2}{2} \log\left(1-\frac{r_{ij}^2}{R_0^2}\right)\text{,}
	\end{equation}
	with $K=100$ and $R_0=1.5$.\par
	Analogously, propulsion directions of dimers also undergo free rotational diffusion without any torques acting on them. Still, the propulsion directions of the bonded pairs could be correlated. The two extreme cases shown schematically in Fig. \ref{fig:schematicDimers} are considered here, namely completely independent propulsion directions (uncorrelated dimers) and a shared direction vector that undergoes the same rotational diffusion (correlated dimers). Both propulsion mechanisms are quite different from the active dimer systems examined by \citet{Cugliandolo:2015} where the propulsion always points along the molecular axis.\par
	\section{Finite size transitions and numerical methods\label{sec:numericalResults}}
	\begin{figure*}[tb]
		\centering
		$\underrightarrow{\hspace{1cm}\text{Packing Fraction}\hspace{1cm}}$\\
		\centering
		\begin{minipage}{0.15\linewidth}
			\includegraphics[width=\linewidth]{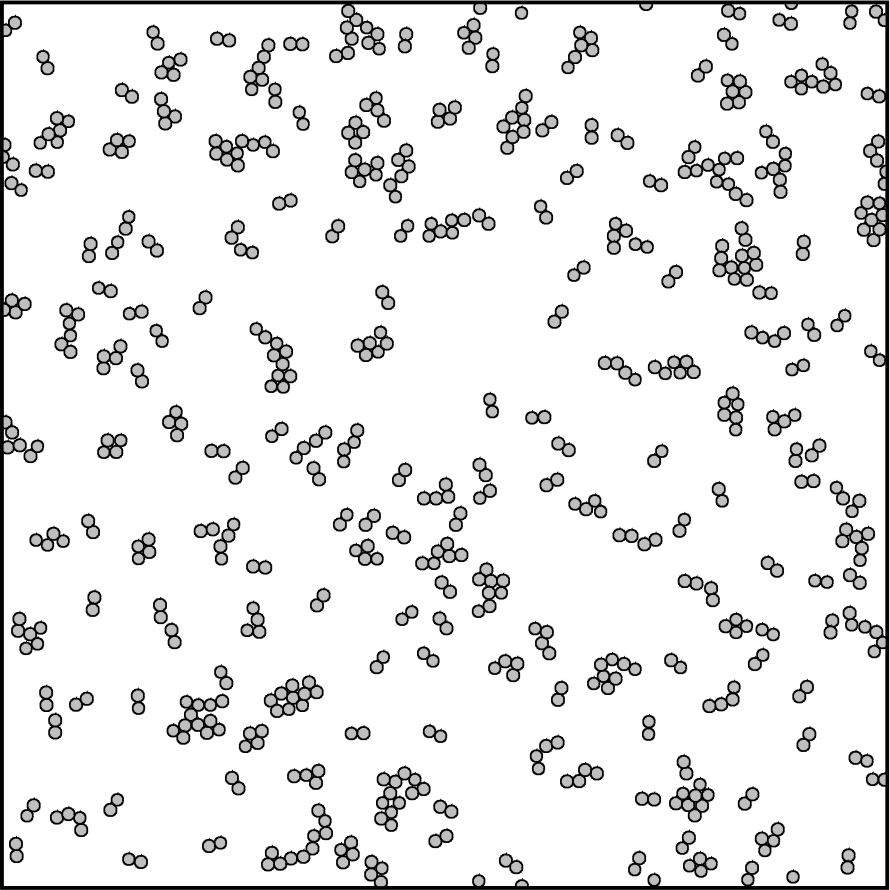}\\
			\centering $\eta=0.1$
		\end{minipage}\hfill
		\begin{minipage}{0.15\linewidth}
			\includegraphics[width=\linewidth]{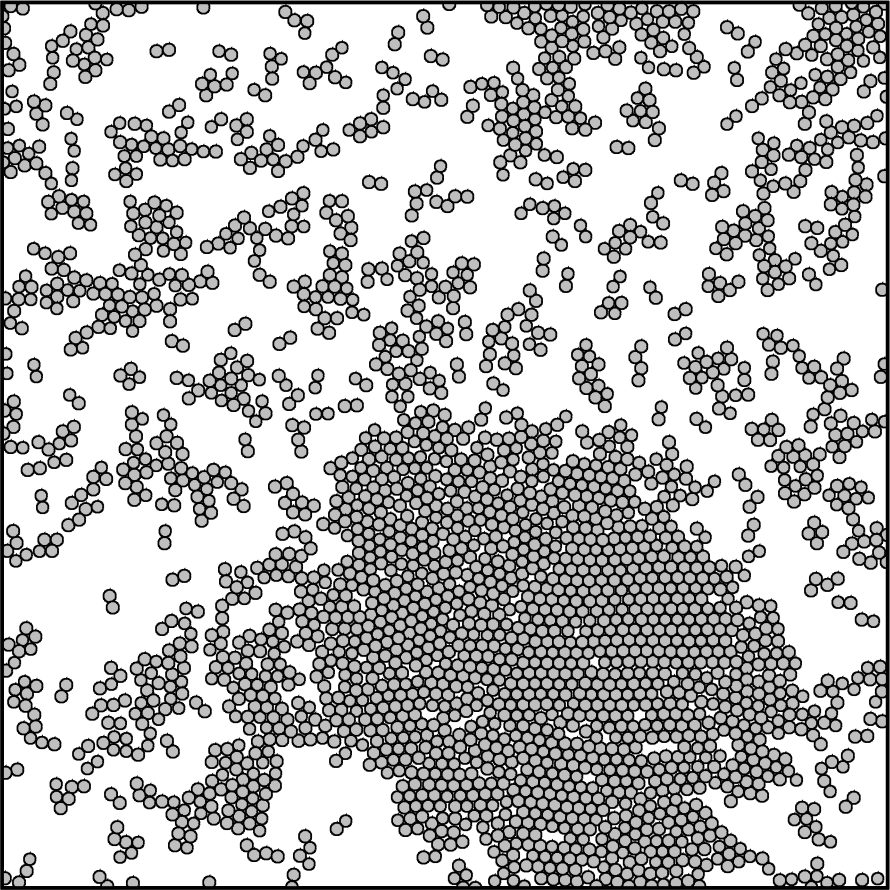}\\
			\centering $\eta=0.4$
		\end{minipage}\hfill
		\begin{minipage}{0.15\linewidth}
			\includegraphics[width=\linewidth]{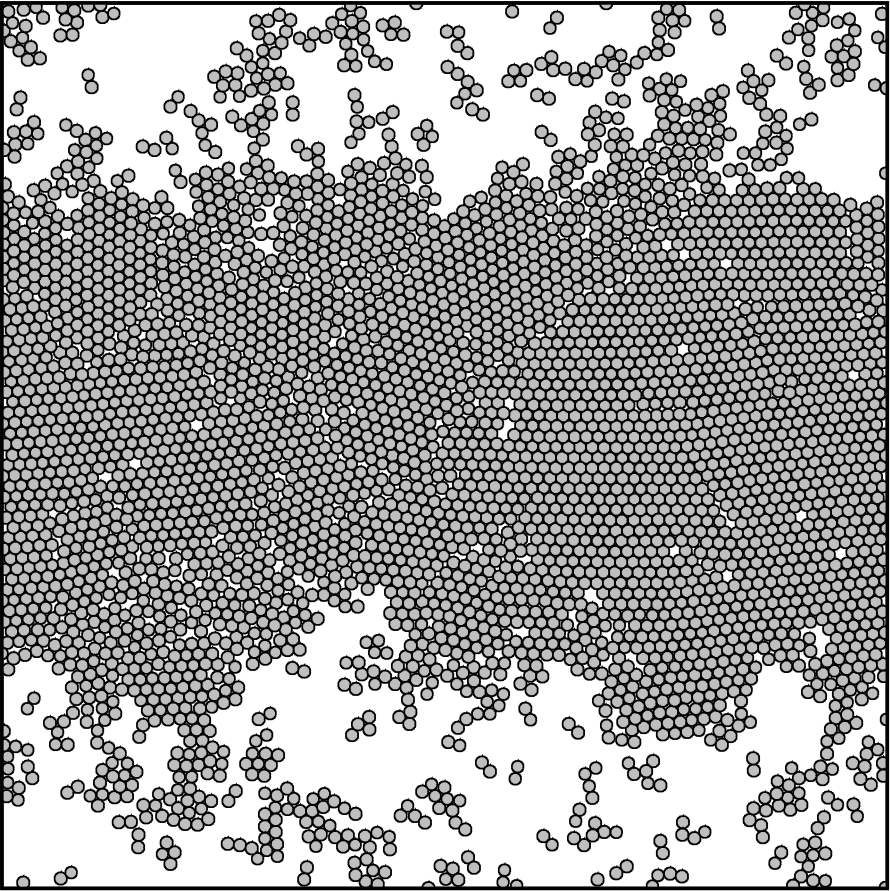}\\
			\centering $\eta=0.6$
		\end{minipage}\hfill
		\begin{minipage}{0.15\linewidth}
			\includegraphics[width=\linewidth]{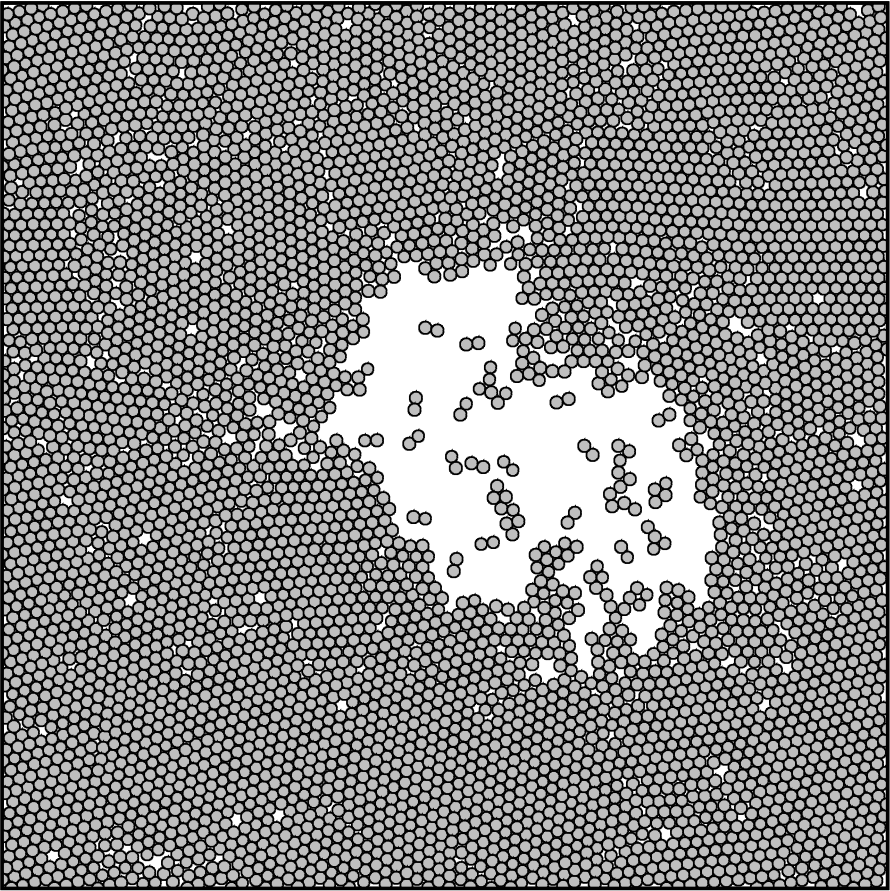}\\
			\centering $\eta=0.8$
		\end{minipage}\hfill
		\begin{minipage}{0.15\linewidth}
			\includegraphics[width=\linewidth]{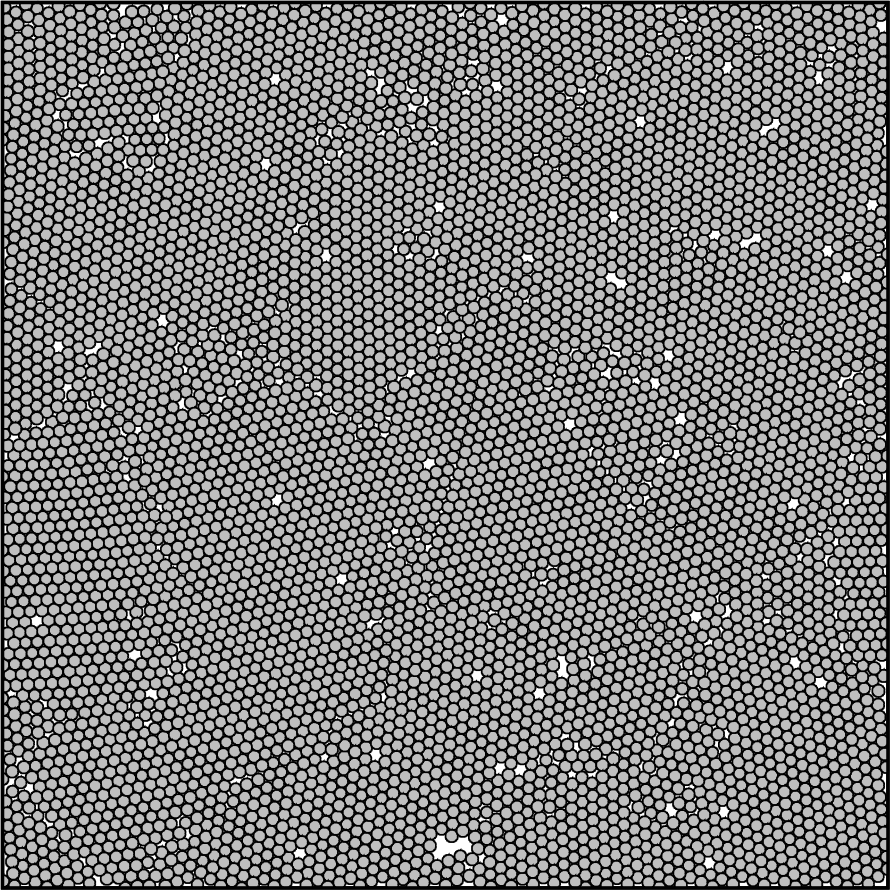}\\
			\centering $\eta=0.91$
		\end{minipage}
		\caption{Finite size transitions of dimer systems in the two phase region ($\text{Pe}\approx188$): For increasing packing fraction at constant propulsion velocity, systems undergo finite size transitions similar to finite systems undergoing a gas-liquid phase transition in equilibrium. For low densities a homogeneous, gas-like phase is stable. Increasing the packing fraction eventually leads to the formation of a liquid-like droplet. For densities well in the phase separated region, a slab configuration is reached. Further increase of the density results in a dilute bubble in a homogeneous dense phase. This bubble will disappear if the packing fraction is high enough.}
		\label{fig:finiteSizeTransitions}
	\end{figure*}
	Above a critical velocity and for intermediate area fractions $\eta=\frac{N\pi d_{\text{BH}}^2}{4A}$  active Brownian monomers as well as dimers undergo phase separation similar to a gas-liquid phase transition in equilibrium. The Peclet number $\text{Pe}=\frac{3v_0\tau_r}{d_{\text{BH}}}$ takes the role of inverse temperature. Varying the packing fraction for fixed propulsion $v_0$, the system shows finite-size transitions as already shown for monomers\cite{Bialke:2015a,Stenhammar:2014} and which have already been studied in detail for equilibrium system (see e.g. \cite{MacDowell:2004,Winter:2009,Schrader:2009,Binder:2012}). The same effect can be observed for active Brownian dimers. At low packing fractions the system will be in a homogeneous gas state. For increasing packing fraction, a droplet of dense active liquid forms. When the droplet's size increases and connects through the periodic boundaries, a slab geometry is reached. After further increase of packing fraction, a bubble of gas in an active liquid is formed that shrinks until only the dense phase is left. This dense phase shows grains of local hexagonal order. But those grains do not represent a true crystal as they are still very dynamic as they move and turn throughout the simulation. They show no sign of merging into a large static crystal during our simulations. Exemplary snapshots of the different finite-size phases for dimer systems are shown in Figure \ref{fig:finiteSizeTransitions}.\par
	\begin{figure}[tb]
		\centering
		\includegraphics[width=\linewidth]{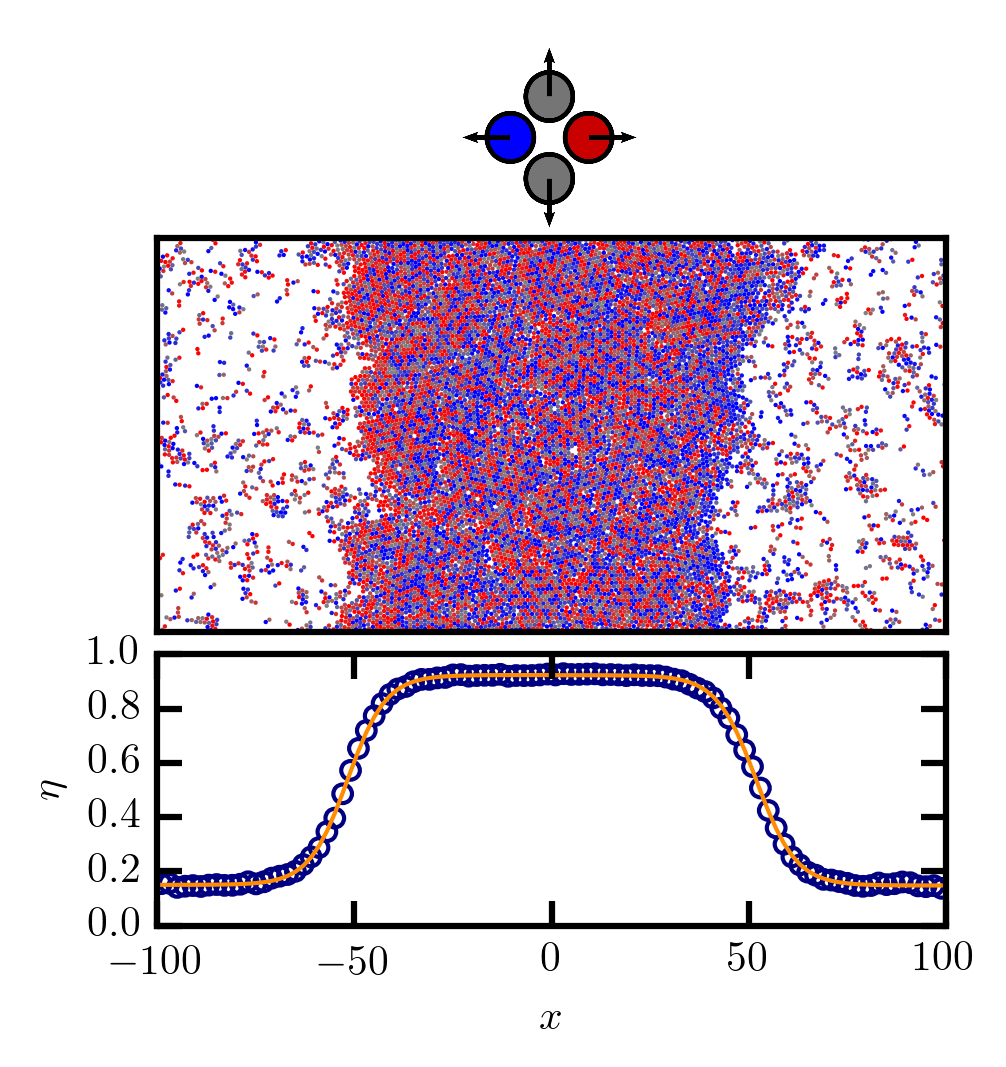}
		\caption{\emph{Top:} Example snapshot of a slab configuration of uncorrelated dimers in an elongated box: Particles are colored according to their orientation in $x$-direction ranging from red for particles pointing in the positive $x$-direction over gray for those aligned with the $y$-axis to blue for particles with a direction anti-parallel to the $x$-axis. Note the polarization of the particles at the interface, where particles are pointing inwards and thus stabilize the slab. \emph{Bottom:} Average density distribution along the $x$-axis: The blue circles are results from simulation. The orange lines show a fit with a hyperbolic tangent (see Eq. (\ref{eq:tanhFit})) for both sides, independently. A plateau forms for both the liquid- and gas-like phases, thus allowing to reliably extract the binodal densities.}
		\label{fig:densityDistribution}
	\end{figure}
	To determine the binodal lines in the phase diagram, we utilize the slab geometry \cite{Das:2014,Bialke:2015a}. Points on the binodal can be found by extraction of coexisting densities for different propulsion strengths. The straightness of the interface in that case minimizes finite-size effects. Simulations in a box elongated along the $x$-axis ensure that the slab aligns perpendicular to that axis \cite{Watanabe:2012,Das:2014,Bialke:2015a}. The binodal densities can be found as plateau values of the density distribution along the elongated axis, which is found by separating the system into bins perpendicular to that axis. The slab is still free to move along the $x$-axis of the box. Therefore, before doing a time average, the center of mass of the system is shifted to zero. One exemplary distribution for active dumbbells is shown in Figure \ref{fig:densityDistribution}. The fit of the form:
	\begin{equation}
		\eta(x) = \frac{\eta_{\mathrm{liq}}+\eta_{\mathrm{gas}}}{2} + \frac{\eta_{\mathrm{liq}}-\eta_{\mathrm{gas}}}{2} \tanh\left(\frac{x-x_0}{2\omega}\right)
		\label{eq:tanhFit}
	\end{equation}
	reproduces the distribution for both sides separately. Here, $\eta_{\mathrm{liq}}$ and $\eta_{\mathrm{gas}}$ are the bulk packing fractions of the liquid and gas phase respectively, while $\omega$ is a measure of the width of the interface. Multiple independent runs allow to estimate the uncertainty of the measurement.\par
	The results for two-dimensional monomers\cite{Bialke:2015a} are shown in Figure \ref{fig:phaseDiagramMonomersDimers}. For all plots without errorbars, errors are smaller than the symbol sizes. Although the method works well and reliably for velocities in this region, the area around the critical point has to be excluded from the analysis as the interface region grows due to the divergent correlation length and thus no well-defined plateau is reached for feasible box sizes.\par
	\section{Active dumbbells}
	\subsection{Phase diagram}
	\begin{figure*}[!tb]
		\centering
		\includegraphics[width=\linewidth]{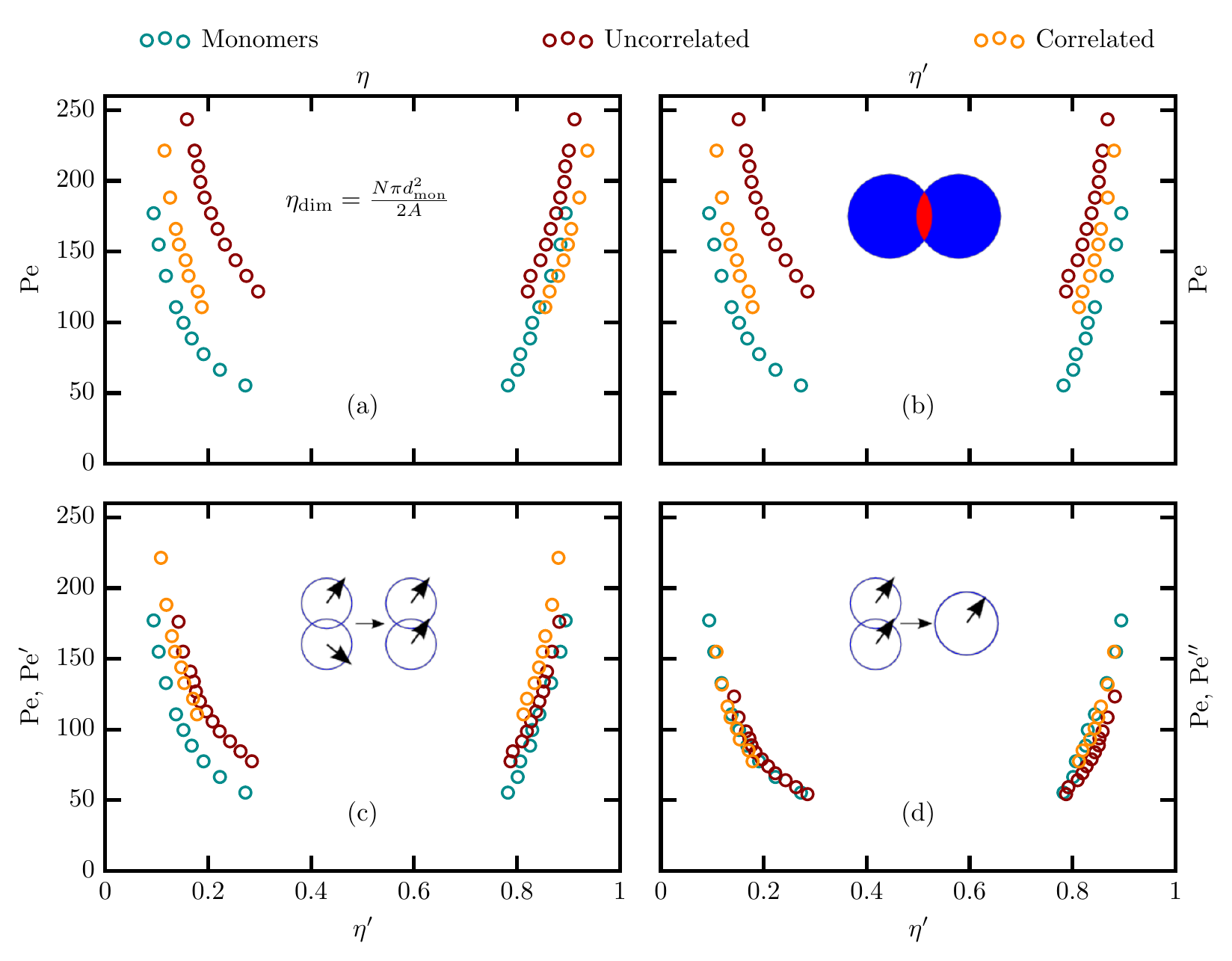}
		\caption{(a) Phase diagrams for active Brownian disks (monomers) and dimers in two dimensions: In blue the binodal lines for active Brownian particles are shown.\cite{Bialke:2015a} Red points indicate the coexistence densities for pairs of active Brownian particles bound by a strong FENE bond. Due to their uncorrelated orientation, much of their propulsion energy is used to stretch the bond, leading to an effectively reduced self-propulsion. Therefore, these dimers phase-separate only for much higher propulsion velocities. In orange, active dimers with shared orientations are shown. Due to alignment, the propulsion energy is not dissipated by the bond stretching. Thus, the correlated dimers start to phase-separate much earlier. (b) Due to overlap, a dimer does not have the same area as two disks. Correcting for this by use of the average bond length for the respective simulation leads to a better agreement of the limiting binodal densities. (c) Uncorrelated and correlated dimers can be mapped by assuming that the velocity of the uncorrelated dimers is effectively the average of the two independent propulsion velocities. (d) By scaling both dimer velocities with a constant factor of $0.7$ to account for additional effects such as different cross-sectional areas and rotations of the dimers they can be mapped onto the monomer curve.}
		\label{fig:phaseDiagramMonomersDimers}
	\end{figure*}
	To check the influence of further interactions and of non-spherical shapes, two dimensional systems of active dimers were considered. As explained in section \ref{sec:model}, two different activity models with either totally uncorrelated or correlated propulsion directions were studied. The resulting phase diagram is shown in Figure \ref{fig:phaseDiagramMonomersDimers}(a). Here, the area covered by a dimer is simply approximated as the area of two non-overlapping disks. To account for possible overlap of dimers due to the additional bond potential, in Figure \ref{fig:phaseDiagramMonomersDimers}(b) the packing fractions of  dimers are scaled by computing the average bond length and thus the average area covered by one dimer.\par
	For both types of activities that were examined in our simulations, dimers phase separate only for much higher propulsion strengths than monomers. This does not contradict earlier research by \citet{Suma:2014}, even though they found that their type of dimers actually phase separates much earlier than a corresponding monomeric system. The difference can be understood, comparing the rotational dynamics of the propulsion directions. While directions of dimers in this paper still undergo free rotational diffusion, either independently or in a synchronized fashion, the active dumbbells in the earlier study where always propelled along their main axis. This leads to a very different rotational motion. Especially inside the dense cluster and also at its surface, dimers will get stuck leading to effectively zero rotational diffusion \cite{Cugliandolo:2015,Cugliandolo:2015a} and thus facilitated phase separation.\par
	Connecting more particles by bonds leading to trimers and chains of increasing length will further increase these effects and thus further suppress phase separation. For sufficiently long chains, the phase separated region should not be reached at reasonable propulsion strengths. In the next section, we present a simple mapping to discuss the differences between uncorrelated and correlated dimers and monomers.
	\subsection{Mapping to active disks}
	To explain the differences between correlated and uncorrelated dimers, a simple argument can be made: In the case of uncorrelated dimers, the propulsion strength is greatly diminished due to bonded particles pulling against each other and thus stretching the bond rather than to self-propel the full dimer. Neglecting all influences of bond flexibility and torques acting on the dimer, an effective propulsion $v_0'$ of dimers can be computed in the infinitely dilute limit as:
	\begin{equation}
	\begin{aligned}
	\dot{\bm{r}}_{\mathrm{com}}&=\frac{\dot{\bm{r}}_1+\dot{\bm{r}}_2}{2}\\
	&=\frac{v_0}{2}\begin{pmatrix}\cos\phi_1+\cos\phi_2\\
	\sin\phi_1+\sin\phi_2\end{pmatrix} + \sqrt{2D}\frac{\bm{R}_{\mathrm{t},1}+\bm{R}_{\mathrm{t},2}}{2}\\
	&=v_0\cos\frac{\phi_1-\phi_2}{2}\begin{pmatrix}\cos\frac{\phi_1+\phi_2}{2}\\
	\sin\frac{\phi_1+\phi_2}{2}\end{pmatrix} + \sqrt{D}\bm{R}_{\mathrm{t}}\text{.}
	\end{aligned}
	\end{equation}
	In this equation, the difference between correlated and uncorrelated dimers becomes evident. Both particles in a correlated dimer share a propulsion direction which undergoes the same rotational diffusion as a monomeric particle. Therefore, the equation of motion for the center of mass reduces to that of a monomer:
	\begin{equation}
	\dot{\bm{r}}_{\mathrm{com, corr}} = v_0 \begin{pmatrix}\cos\phi\\
	\sin\phi\end{pmatrix} + \sqrt{D}\bm{R}_{\mathrm{t}}\text{.}
	\end{equation}
	The propulsion directions of the two particles forming an uncorrelated dimer are independent. Therefore, we can change variables to $\bar{\phi}=\frac{\phi_1+\phi_2}{2}$ and $\tilde{\phi}=\frac{\phi_1-\phi_2}{2}$ and end up with an equation of motion of the form:
	\begin{equation}
	\begin{aligned}
	\dot{\bm{r}}_{\mathrm{com, uncorr}}&=v_0\cos\tilde{\phi}\begin{pmatrix}\cos\bar{\phi}\\
	\sin\bar{\phi}\end{pmatrix} + \sqrt{D}\bm{R}_{\mathrm{t}}\\
	\dot{\bar{\phi}}&=\sqrt{D_{\mathrm{r}}}R_{\mathrm{r, 1}}\\
	\dot{\tilde{\phi}}&=\sqrt{D_{\mathrm{r}}}R_{\mathrm{r, 2}}\text{.}
	\end{aligned}
	\end{equation}
	The variable $2\tilde{\phi}$ is the sum of two random variables $\phi_1$ and $-\phi_2$ that are uniformly distributed in $[-\pi, \pi)$ and as such its probability distribution is given as the convolution of a uniform distribution with itself:
	\begin{equation}
	p_{2\tilde{\phi}}(2\tilde{\phi}) = \int_{-\infty}^{\infty}\d{\phi} p_{\phi}(\tilde{\phi}-\phi) p_{\phi}(\phi)\text{,}
	\end{equation}
	where:
	\begin{equation}
	p_{2\phi}(2\phi)=\begin{cases}\frac{1}{2\pi}& \phi\in[-\pi,\pi)\\0& \text{otherwise}\end{cases}\text{.}
	\end{equation}
	Therefore, $p_{2\tilde{\phi}}(2\tilde{\phi})$ can be rewritten using the Heaviside step function $\Theta(x)$:
	\begin{equation}
	\begin{aligned}
	p_{2\tilde{\phi}}(2\tilde{\phi}) &=\frac{1}{2\pi}\int_{-\infty}^{\infty}\d{\phi}
	\begin{array}[t]{l}
		\Theta(\pi+\phi-2\tilde{\phi})\cdot\Theta(\pi-\phi+2\tilde{\phi})\cdot\\[.15cm]
		\Theta(\pi+\phi)\cdot\Theta(\pi-\phi)
	\end{array}\\
	&= \frac{1}{2\pi}\left(\min(2\tilde{\phi}+\pi, \pi)-\max(2\tilde{\phi}-\pi, -\pi)\right)\\
	&=
	\begin{cases}
	\frac{1}{2\pi}+\frac{2\tilde{\phi}}{4\pi^2}& 0 > 2\tilde{\phi} > -2\pi\\
	\frac{1}{2\pi}-\frac{2\tilde{\phi}}{4\pi^2}& 0 <  2\tilde{\phi} < 2\pi
	\end{cases}\text{.}
	\end{aligned}
	\end{equation}
	Thus, the probability distribution of $\tilde{\phi}$ is given by:
	\begin{equation}
	p_{\tilde{\phi}}(\tilde{\phi}) =
	\begin{cases}
	\frac{1}{\pi}+\frac{\tilde{\phi}}{\pi^2}& 0 > \tilde{\phi} > -\pi\\
	\frac{1}{\pi}-\frac{\tilde{\phi}}{\pi^2}& 0 < \tilde{\phi} < \pi
	\end{cases}\text{.}
	\end{equation}
	Effectively, the propulsion speed is thus decreased by a factor of:
	\begin{equation}
	\begin{aligned}
	\left\langle|\cos\tilde{\phi}|\right\rangle &= \int_{-\pi}^{\pi}\d{\tilde{\phi}}p(\tilde{\phi})|\cos\tilde{\phi}|\\
	&= 2\int_{0}^{\pi}\d{\tilde{\phi}}p(\tilde{\phi})|\cos\tilde{\phi}|\\
	&= \frac{2}{\pi}
	\end{aligned}
	\end{equation}
	Rescaling the velocities of the uncorrelated dimers with this factor results in Figure \ref{fig:phaseDiagramMonomersDimers}(c). Here, uncorrelated and correlated dimers match remarkably well, indicating that this is indeed the main difference of those two systems. Still, dimers have a different cross sectional area and interact with surrounding particles differently.\par
	Scaling the velocity of both types of dimers by another factor of approximately $0.7$, a rather good agreement of all curves can be reached, as can be seen in Figure \ref{fig:phaseDiagramMonomersDimers}(d). This heuristic factor includes all contributions of shape, anisotropy, cross sectional area, and rotation of the full dimer. The mapping works especially well for correlated dimers. For uncorrelated dimers, the curves do not match as well. These differences stem from additional effects of the non-uniform effective swimming speed and the increase of relative motion of the particles forming a dimer.\par
	\section{Active spheres in three dimensions}
\begin{figure}[tb]
		\centering
		\includegraphics[width=\linewidth]{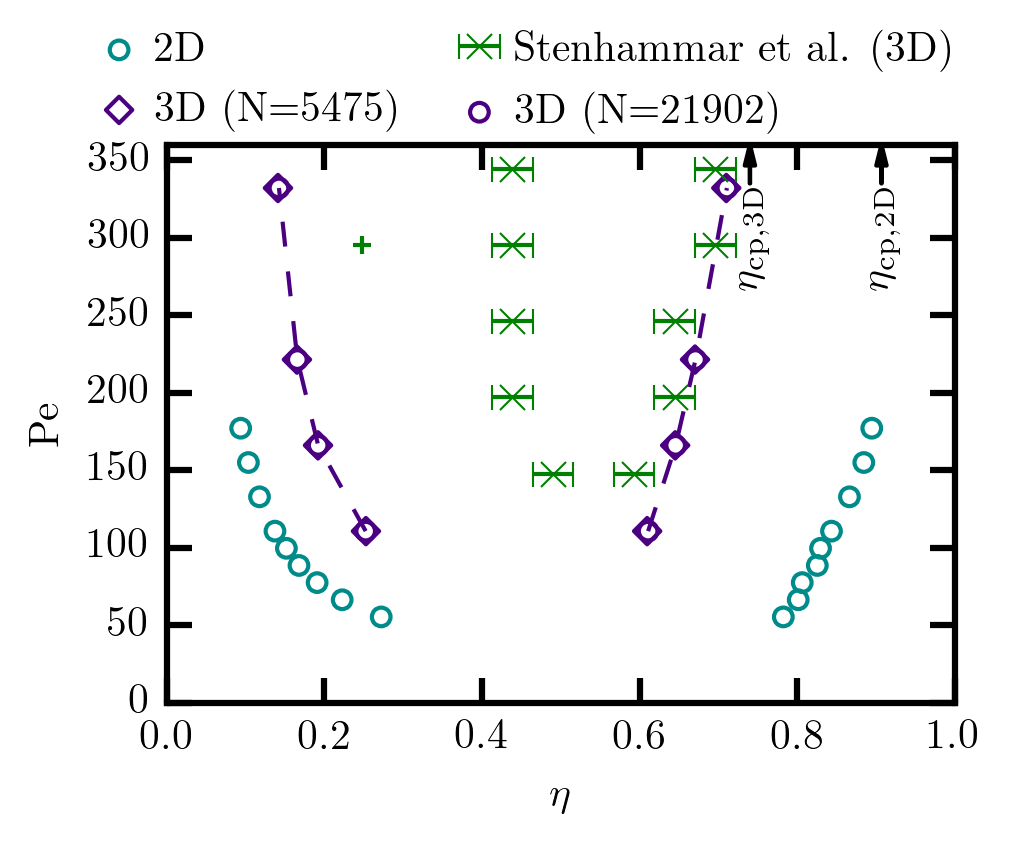}
		\caption{Phase diagrams for active Brownian particles in two and three dimensions: Here, $\eta$ represents area and volume fraction in 2D and 3D, respectively. In blue the binodal lines for active Brownian particles in two dimensions are shown\cite{Bialke:2015a}. Purple points indicate the coexistence densities for three dimensional particles. The connecting lines are included as guide for the eye. Results for a smaller system are marked by purple diamonds. In all cases the errors are smaller than the symbol size. For comparison the phase diagram reported by \citet{Stenhammar:2014} is shown as green crosses. Their estimate for a point on the dilute binodal is shown as a green plus.}
		\label{fig:phaseDiagram2D3D}
	\end{figure}

	Active Brownian particles in three dimensions also show phase separation\cite{Stenhammar:2014,Wysocki:2014}. To determine the phase diagram, we employ the same method described in section \ref{sec:model}. We placed $N=21902$ spheres in a box with side lengths $L_{\mathrm{x}} = 2.5L_{\mathrm{y}} = 2.5L_{\mathrm{z}}=60$ and simulated for different propulsion velocities. Earlier work by \citet{Stenhammar:2014} and \citet{Wysocki:2014} already predicted that higher velocities are needed for phase separation in 3D. This is verified by our quantitative and finite-size independent analysis of the phase diagram. Results are shown in Figure \ref{fig:phaseDiagram2D3D}. Note that in three dimensions $\eta$ refers to volume fraction rather than an area fraction that is considered in the two dimensional case. Thus, a comparison of the width of phase separated regions is not meaningful. Nonetheless, one can see that phase separation only occurs at much higher propulsion velocities with $\text{Pe}_{0,3D}\gtrsim80$ whereas in two dimensions it sets in already for lower velocities corresponding to $\text{Pe}_{0,2D}\gtrsim40$.\par
	This can be understood from a simple kinetic argument. There are two time scales that govern the system, namely the reorientation time $\tau_{\text{r}}$ and the mean collision time $\tau_{\text{c}}$, which is related to the mean free path by the active propulsion $v_0$\cite{Redner:2013,Stenhammar:2013}. The ratio between those time scales determines, whether spheres likely will get stuck on collision leading to phase separation. Since for a system with rotational diffusion coefficient $D_{\text{r}}$, the reorientation in two dimensions $\tau_{\text{r,2D}}=D_{\text{r}}^{-1}$ is twice as large as in three dimensions $\tau_{\text{r,3D}}=(2D)_{\text{r}}^{-1}$, smaller velocities and thus smaller mean collision times already lead to phase separation in two dimensions, whereas for a three dimensional system, much higher propulsion strengths are needed.\par
	To verify that our results are indeed not finite-size dependent, we simulated a second smaller system with $L_{\mathrm{x}} = 5L'_{\mathrm{y}} = 5L'_{\mathrm{z}}$ and $N'=5475$ particles. The two curves nicely fall onto each other, as indicated in Figure \ref{fig:phaseDiagram2D3D}. In the same figure, the phase diagram reported by \citet{Stenhammar:2014} is shown. In contrast to our study, they used $\sigma$ as characteristic length scale. To allow for comparison with our results, their data were scaled by use of the appropriate Barker-Henderson diameter\cite{Barker:1967} $d_{\mathrm{BH}} \approx 1.01561$. The difference in the diameter stems from their use of a softer WCA-potential with $\epsilon=1$. Also they are varying $\text{Pe}$ by changing $\kT$ and thus $\tau_{\text{r}}$ rather than $v_0$.\par
	In their study, the phase diagram was scanned for spontaneous nucleation during a fixed simulation time frame, effectively determining an evaporation-condensation type transition for a liquid droplet, which the authors refer to as the spinodal. Note that these events are finite-size dependent in the passive case\cite{MacDowell:2004,Winter:2009,Schrader:2009,Binder:2012} and the corresponding lines lie within the binodal lines. A similar trend can be seen in another study by \citet{Wysocki:2014} that also reports values for the phase boundaries well within our binodal lines. They introduce the activity similar to our current study but used a Yukawa type potential rather than the more short-ranged WCA-potential.\par
	By starting from a dense droplet and investigating its stability at $\text{Pe}\approx295$ and for different packing fractions, \citet{Stenhammar:2014} also report one point on the dilute binodal. This is also measuring the droplet transition and thus this point should be strongly finite-size dependent, as well. As they examined a very large system, the large discrepancy between this point and the binodal line determined in this report is surprising and requires further inspection.\par
	\section{Conclusions\label{sec:conclusions}}
	In this paper, we provided high precision estimates for binodal lines in three models of phase separating active Brownian particles. Influences of dimensionality as well as anisotropic shapes in the form of active Brownian dumbbells have been studied and compared to previous results for active Brownian disks \cite{Bialke:2015a}.\par
	The minimum Peclet number necessary for phase separation of active spheres in three dimensions is higher than that of disks in two dimensions. We provided values for the phase boundaries which are finite-size independent and thus can serve as reference values for the binodal lines of active Brownian spheres.\par
	Active Brownian dumbbells also phase separate but at even higher Peclet numbers. Here, the influence of correlations of the propulsion directions plays an important role. To understand its influence, the extreme cases of fully correlated, and totally uncorrelated dimers were studied. While correlated dumbbells with a shared director already phase separate at higher propulsion strengths compared to corresponding monomers, for uncorrelated dimers even higher propulsion velocities are needed. This can be understood in terms of an effectively reduced propulsion velocity in the case of uncorrelated motion, where the particles can also pull against each other.
	\section*{Acknowledgements}
	JTS, TS, and PV gratefully acknowledge financial support by	DFG within priority program SPP 1726 (Grants No. SP 1382/3-1 and VI 237/5-1).  ZDV Mainz is acknowledged for computing time on the MOGON supercomputer.
	\bibliography{bondPaper.bib}
\end{document}